\documentclass[12pt]{iopart}
\usepackage{iopams}
\usepackage{epsfig}
\newcommand{\be}{\begin{equation}}
\newcommand{\ee}{\end{equation}}
\newcommand{\bd}{\begin{displaymath}}
\newcommand{\ed}{\end{displaymath}}
\newcommand{\vsp}{\vspace*{3mm}}

\newcommand{\bra}{\langle}
\newcommand{\ket}{\rangle}
\newcommand{\order}{{\cal O}}

\newcommand{\sgn}{{\rm sgn}}

\newcommand{\id}{{\rm 1\!\!I}}

\newcommand{\bq}{\ensuremath{\mathbf{q}}}

\newcommand{\bR}{\ensuremath{\mathbf{R}}}

\newcommand{\boldxi}{{\mbox{\boldmath $\xi$}}}
\newcommand{\boldomega}{{\mbox{\boldmath $\omega$}}}

\newcommand{\boldOmega}{{\mbox{\boldmath $\Omega$}}}

\begin{document}

\title[Dynamics of a spherical minority game]{Dynamics of a spherical minority game}

\author{T Galla\dag, A C C Coolen\ddag\ and D Sherrington\dag
%\footnote[3]{To
%whom correspondence should be addressed (romneya.robertson@iop.org)}
}

\address{\dag\ Department of Physics, University of Oxford,
Theoretical Physics, 1 Keble Road, Oxford OX1 3NP, UK}

\address{\ddag\ Department of Mathematics, King's College London, The Strand,
London WC2R 2LS, UK}

\begin{abstract}
We present an exact dynamical solution of a spherical version of
the batch minority game (MG) with random external information. The
control parameters in this model are the  ratio of the number of
possible values for the public information over the number of
agents, and the radius of the spherical constraint on the
microscopic degrees of freedom. We find a phase diagram with three
 phases: two without anomalous response (an oscillating
versus a frozen state), and a further frozen phase with divergent
integrated response. In contrast to standard MG versions, we can
also calculate the volatility exactly. Our study reveals
similarities between the spherical and the conventional MG, but
also intriguing differences. Numerical simulations confirm our
analytical results.
\end{abstract}


\ead{\tt galla@thphys.ox.ac.uk, tcoolen@mth.kcl.ac.uk,
d.sherrington1@physics.oxford.ac.uk}

\section{Introduction}

The dynamics of interacting agents is currently studied intensively,
applying the ideas and techniques of equilibrium and non-equilibrium
statistical mechanics. One of the models which has attracted
particular attention is the so-called minority game (MG), introduced
as a minimalist econophysics model for a financial market
\cite{ChalZhan97}. The players in the MG are traders who, at each
round of the game, have to make one of two possible choices (e.g. buy
or sell) in response to publicly available information. Each aims to
make profit by making the opposite choice to the majority of agents.
The interaction between agents is indirect: they cannot observe
individual actions of others, but only the subsequent cumulative
effect of all actions on the market. To determine their own trading
actions, each agent holds a pool of strategies, assigned randomly
before the start of the game and then kept fixed. These effectively
act as look-up tables, mapping the observation of publicly available
information onto a proposed trading action. In the versions of the MG
studied so far (e.g.
\cite{ChalZhan97,CavaGarrGiarSher99,ChalMarsZecc00,HeimCool01}) agents
cannot combine strategies, but select the one which they regard as
their best. We refer to those types of MGs as conventional. The
identification of the best strategy is based on points the agents
allocate to each of their strategies in order to measure their
performance. After each round of the game each agent evaluates the
quality of each of his or her strategies, increasing the points of
strategies which would have yielded a correct minority prediction. For
a general overview of MG-type models we refer to \cite{Chalweb}.

The update rules of the MG look simple, but describe surprisingly
complex cooperative processes. This is most visible in the non-trivial
behaviour of the fluctuations of the total bid, the so-called market
volatility \cite{ChalZhan97, Cava99,
SaviManuRiol99,CavaGarrGiarSher99, ChalMarsZecc00}. The main control
parameter in MGs is the ratio $\alpha=p/N$ of the number $p$ of
possible values for the public information over the number $N$ of
players. One observes a critical value $\alpha_c$ which marks a
dynamical\footnote{MGs do not obey detailed balance, so one can only
speak about non-equilibrium phase transitions.} phase transition,
separating a non-ergodic phase ($\alpha<\alpha_c$) from an ergodic one
($\alpha>\alpha_c$). In the non-ergodic phase, the volatility is very
sensitive to initial conditions
\cite{ChalMarsZecc00,MarsChalZecc00,GarrMoroSher00}, and the
integrated response is infinite \cite{HeimCool01}. Moreover, in the
stationary state the system exhibits persistent oscillations in the
non-ergodic phase, whereas oscillations decay on finite time scales
for $\alpha>\alpha_c$ \cite{HeimCool01, SherGall03}.

Analytical progress is possible using equilibrium and dynamical
approaches and has resulted in analytical expressions for $\alpha_c$,
which are now regarded as exact \cite{ChalMarsZhan00,
HeimCool01,CoolHeim01,CoolHeimSher01}. The generating functional
analysis \`a la De Dominicis \cite{DeDom78} has proven particularly
valuable; it enabled a full understanding of the dynamics of the MG in
the ergodic phase. In this formalism, the strategy selection dynamics
of the agents is mapped onto a non-Markovian effective single agent
process. In the case of conventional MGs, the microscopic laws and the
resulting single-trader process are non-linear and resist analytical
solution. Instead, one derives a coupled set of implicit equations for
stationary states, from which one tries to extract the values of the
persistent order parameters. In contrast to equilibrium systems, there
are no fluctuation-dissipation relations which could be used to
simplify those equations. In the ergodic phase the analysis can be
simplified taking into account the existence of so-called `frozen
agents' (runaway solutions of the microscopic laws). A proper
understanding of the dynamics in the non-ergodic regime, however, is
still lacking. Moreover, for conventional MGs the market volatility
(the MG's main observable) cannot be expressed in terms of persistent
order parameters. Instead, detailed knowledge of both the long-time
and the short-time behaviour of the macroscopic order parameters is
required. Hence, even in the ergodic phase, results for the volatility
are so far restricted to approximations, whereas in the non-ergodic
phase only approximate asymptotic results in the limit $\alpha\to0 $
are available \cite{HeimCool01}.  For a recent review of dynamical MG
analyses see e.g. \cite{Coolenreview}. Approximations for the
volatility in the ergodic state are also accessible within the
framework of replica theory \cite{ChalMarsZecc00}.

In this paper we present a version of the MG which is analytically
solvable, but nevertheless displays some of the interesting features
found in the conventional MG. To this end we study the dynamics of a
spherical version of the MG using the generating functional
approach. Like in spherical $p$-spin glasses with polynomial equations
of motion for the continuous microscopic degrees of freedom, {\em
explicit} closed equations for the two-time correlation and response
functions can be formulated \cite{CrisHornSomm}.

A second control parameter, the radius $r$ of the sphere to which the
dynamics is confined, becomes relevant in the present model. Apart
from the spherical constraint we choose the update rules to be linear
in this paper, so that we can solve the resulting dynamical equations
exactly, reminiscient of the $p=2$ case known for spherical
spin-glasses \cite{CuglDean95}. In particular we are able to compute
the volatility in all regions of the phase diagram without making any
approximations at any stage. In terms of the decision making of the
individual agents the linear spherical model corresponds to allowing
them to play linear combinations of strategies (rather than to pick
the best one).

Despite its simple microscopic rules, the spherical MG as presented in
this paper shows interesting behaviour and exhibits novel features as
well as properties analogous to the ones of conventional MGs. In
particular we find three distinct phases in the
$(\alpha,r)$-plane. Our analytical findings are verified convincingly
by numerical simulations.

\section{Model Definitions}

Before defining the spherical version of the MG and giving an
interpretation of its update rules, we will recall the dynamical rules
of a conventional MG as studied for example in \cite{HeimCool01,
CoolHeim01}. We label the $N$ agents in the MG with Roman indices. At
each round $t$ of the game each agent $i$ takes a trading decision
$b_i(t)\in{\rm I\!R}$ (a `bid') in response to the observation of
public information $I_{\mu(t)}$ which is chosen randomly and
independently from a set with $p=\alpha N$ possible
values\footnote{This is the so-called MG with random external
information; in the early MG definition \cite{ChalZhan98} the external
information was not random but coded for the actual history of the
global market.}, so $\mu(t)\in\{1,\dots,\alpha N\}$. The rescaled
total market bid at round $t$ is defined as $A(t)=N^{-1/2}\sum_i
b_i(t)$. Each agent $i$ has $S\geq 2$ fixed trading strategies
(look-up tables) $\bR_{ia}=(R_{ia}^1,\dots,R_{ia}^{\alpha N})$ at his
or her disposal, with $a=1,\dots,S$.  If agent $i$ decides to use
strategy $a$ in round $t$ of the game, his or her bid at this stage
will be $b_i(t)=R_{ia}^{\mu(t)}$. All strategies $\bR_{ia}$ are chosen
randomly and independently before the start of the game; they
represent the quenched disorder of this problem. The behaviour of the
MG was found not to depend much on the value of $S$
\cite{ChalZhan98,ManuLiRiolSavi98}, nor on whether bids are discrete
or continuous \cite{CavaGarrGiarSher99}. For convenience, we choose
$S=2$ and $\bR_{ia}\in\{-1,1\}^{\alpha N}$ in this paper. In order to
decide which strategy to use, the agents assign points $p_{ia}(t)$ to
each of their strategies, on the basis of what would have happened if
they had played that particular strategy:
\begin{equation}
p_{ia}(t) = p_{ia}(t) - R_{ia}^{\mu(t)} A(t). \label{eq:scores}
\end{equation}
Strategies which would have produced a minority decision are thus
rewarded. In the conventional MG, at each round $t$ each player $i$
uses the strategy in his or her arsenal with the highest score,
i.e. $b_i(t)=R_{i\tilde a_i(t)}^{\mu(t)}$, where $\tilde a_i(t) =
\mbox{arg max}_a\, p_{ia}(t)$. For $S=2$ the rules (\ref{eq:scores})
can then be simplified upon introducing the differences
$q_i(t)=\frac{1}{2}[p_{i1}(t)-p_{i2}(t)]$. Thus, if $q_i(t)>0$, agent
$i$ plays strategy $\bR_{i1}$, whereas for $q_i(t)<0$ he or she plays
$\bR_{i2}$. Hence, in the conventional MG
$b_i(t)=\omega_i^{\mu(t)}+\sgn[q_i(t)]\xi_i^{\mu(t)}$, where
$\boldomega_i=\frac{1}{2}[\bR_{i1}+\bR_{i2}]$ and
$\boldxi_i=\frac{1}{2}[\bR_{i1}-\bR_{i2}]$. The evolution of the
$\{q_i\}$ is given by
\begin{equation} \label{onlineupdate}
 q_i(t+1) = q_i(t) - \xi_i^{\mu(t)}\left[
\Omega^{\mu(t)}+\frac{1}{\sqrt{N}}\sum_j \xi_j^{\mu(t)}
\mbox{sgn}[q_j(t)]\right]
\end{equation}
with $\boldOmega=N^{-1/2}\sum_j \boldomega_j$. Equation
(\ref{onlineupdate}) defines the standard (or so-called `on-line')
MG. Alternatively, corresponding to updating the $\{q_i\}$ only every
${\cal O}(N)$ time-steps, one might define the dynamics in terms of an
average over all possible values of the external information in
(\ref{onlineupdate}), resulting in the so-called (conventional) `batch'
MG \cite{HeimCool01}:
\begin{equation}\label{batchupdate}
q_i(t+1)=q_i(t)-h_i-\sum_j J_{ij}~ \mbox{sgn}[q_j(t)].
\label{eq:batchMG}
\end{equation}
Here $J_{ij}=2N^{-1}\boldxi_i\cdot\boldxi_j$ and
$h_i=2N^{-1/2}\boldxi_i\cdot\boldOmega$. See \cite{CoolHeimSher01}
for stochastic extensions and \cite{SherGall03} for consideration of
the effects of anti-correlation of strategies on the comparison of
on-line and batch models. \vsp

The batch model (\ref{eq:batchMG}) is particularly suitable for being
replaced by a spherical version. We first linearize
(\ref{eq:batchMG}), and subsequently normalize\footnote{The
spherical normalisation is necessary to suppress possible runaway
solutions corresponding to eigenmodes of the linear update rule with
eigenvalues of a modulus larger than one.} the vector
$(q_1,\ldots,q_N)$ to a fixed length $r>0$ at each iteraction step
$t$, resulting in the spherical batch MG:
\begin{eqnarray}
[1+\lambda(t+1)]q_i(t+1)=q_i(t)-h_i-\sum_j J_{ij}~ q_j(t),
\label{eq:sphericalupdate}
\\[-2mm]
\frac{1}{N}\sum_i q^2_i(t)=r^2~~~~{\rm for~all}~ t.
\label{eq:constraint}
\end{eqnarray}
The values of the constraining forces $\lambda(t)$ in
(\ref{eq:sphericalupdate}) follow from (\ref{eq:constraint}); we
exclude artificial sign changes by insisting on $1+\lambda(t)>0$ for
all $t$. We note that our model
(\ref{eq:sphericalupdate},\ref{eq:constraint}) has no analogue of the
{\em tabula rasa} MG initialization, $q_i(0)=0$ for all $i$, often
employed in the conventional MG.  We also note that the $\{q_i\}$ of
the conventional MG (\ref{eq:batchMG}) do not satisfy a spherical
constraint, unlike the spins of a conventional Ising spin
system. There is thus no reason for restricting oneself to $r=1$. In
fact, $r$ is a new control parameter and the system exhibits phase
behaviour in the $(\alpha,r)$-plane with interesting differences from
the conventional game.

The linearity of (\ref{eq:sphericalupdate}) implies that agents now
play {\em linear combinations} of their strategies. Upon presentation
of public information $I_\mu$ at time $t$ the bid of player $i$ in a
corresponding on-line game is \be
b_i(t)=\frac{1}{2}[1+q_i(t)]R^{\mu}_{i1}+\frac{1}{2}[1-q_i(t)]R^{\mu}_{i2}.
\label{eq:combine} \ee 
The main object of natural interest in
MGs is the volatility, which describes the standard deviation of the total
(re-scaled) market bid \be
A^\mu[\bq(t)]=\frac{1}{\sqrt{N}}\sum_i\left[\omega_i^{\mu}+q_i(t)\xi_i^{\mu}\right].
\label{eq:totalbid} \ee In the on-line models the relevant averages
are over the stochasticity of the `information'. In deterministic
batch problems, such as discussed here, these averages are replaced by
ones over $\mu$: $\bra A_t\ket=p^{-1}\sum_{\mu=1}^p A^\mu[\bq(t)]$ and
$\bra A_t A_{t^\prime}\ket=p^{-1}\sum_{\mu=1}^p
A^\mu[\bq(t)]A^\mu[\bq(t^\prime)]$. The volatility is defined as
$\sigma_t^2=\bra A^2_t\ket-\bra A_t\ket^2$. Here we follow
\cite{HeimCool01} and define a more general object, the volatility
matrix $\Xi_{tt^\prime}=\bra A_t A_{t^\prime}\ket-\bra A_t\ket\bra
A_{t^\prime}\ket $:
\begin{equation}
\hspace{-1cm} \Xi_{tt^\prime}=\frac{1}{p}\sum_{\mu=1}^p
A^\mu[\bq(t)] A^\mu[\bq(t^\prime)]
-\left[\frac{1}{p}\sum_{\mu=1}^p A^\mu[\bq(t)]\right]
\left[\frac{1}{p}\sum_{\mu=1}^p A^\mu[\bq(t^\prime)]\right].
\label{eq:volamatrix}
\end{equation}
Note that $\sigma_t^2 =\Xi_{tt}$. Random trading, with $\bq(t)$ taken
randomly and independently from the sphere $\bq^2(t)=Nr^2$ at each
time $t$, would result in $\bra A_t\ket=\order(N^{-\frac{1}{2}})$ and
$\Xi_{tt^\prime}=\frac{1}{2}+\frac{1}{2}r^2\delta_{tt^\prime}+\order(N^{-\frac{1}{2}})$.
The volatility measures the efficiency of the market, with
$\sigma^2_t=0$ corresponding to a perfect match between supply and
demand at time $t$.

\section{Macroscopic Dynamics}

The similarity between the spherical batch MG
(\ref{eq:sphericalupdate},\ref{eq:constraint}) and the conventional
batch MG (\ref{eq:batchMG}) allows us to obtain the effective
single trader equations for
(\ref{eq:sphericalupdate},\ref{eq:constraint}) simply by making
the substitutions $q(t+ 1)\to [1+\lambda(t+1)]q(t+1)$ and
$\sgn[q(t)]\to q(t)$ in the results  of \cite{HeimCool01} (found
within the generating functional analysis framework, in the limit
$N \to\infty$):
\begin{equation} \label{eq:singleagent}
[1+\lambda(t+ 1)] q(t+ 1) = q(t) + \theta(t) - \alpha
\sum_{t^\prime} (\id+ G)^{-1}_{tt^\prime} q(t^\prime) +
\sqrt{\alpha} ~\eta(t).
\end{equation}
Here $\theta(t)$ is an external perturbation field introduced to
generate response functions and $\eta(t)$ is a zero-average Gaussian
noise, characterized by the following covariance matrix (with
$D_{tt^\prime}=1+ C_{tt^\prime}$ and
$\id_{tt^\prime}=\delta_{tt^\prime}$):
\begin{equation}
\Sigma_{tt^\prime}= \bra \eta(t)\eta(t^\prime)\ket_*  =
\left[(\id+G)^{-1} D (\id+G^T)^{-1}\right]_{tt^\prime}.
\label{eq:covariance}
\end{equation}
 The matrices $C$ and $G$
and the constraining forces $\lambda(t)$ are the dynamical order
parameters of the problem, to be determined self-consistently by
solving
\begin{equation}\label{eq:CGselfconsistent}
C_{tt^\prime}=\bra q(t)q(t^\prime)\ket_*, ~~~~~~~~
G_{tt^\prime}=\frac{\partial}{\partial\theta(t^\prime)}\bra
q(t)\ket_*,~~~~~~~~C_{tt}=r^2.
\end{equation}
One always has $\bra q(t)\ket_*=0$. The physical meaning of $C$ and
$G$ is given by
\begin{eqnarray}
 C_{tt^\prime}&=&\lim_{N\to\infty}\frac{1}{N}\sum_i \overline{\bra
q_i(t)q_i(t^\prime)\ket}, \label{eq:meaningC}\\
G_{tt^\prime}&=&\lim_{N\to\infty}\frac{1}{N}\sum_i
\frac{\partial}{\partial \theta_i(t^\prime)}\overline{\bra
q_i(t)\ket} \label{eq:meaningG},
\end{eqnarray}
where $\overline{\cdots}$ denotes an average over the disorder,
i.e. over the space of all strategies in the context of the MG. The
brackets $\bra \ldots\ket_*$ in (\ref{eq:covariance}) and
(\ref{eq:CGselfconsistent}) refer to averaging over the realisations
of the process (\ref{eq:singleagent}), i.e. over the noise
$\{\eta(t)\}$. As usual, the single-agent process
(\ref{eq:singleagent}) is non-Markovian.

It is possible to convert the system
(\ref{eq:singleagent},\ref{eq:covariance},\ref{eq:CGselfconsistent})
into a pair of explicit iterative equations for $C$ and $G$. An
explicit equation for $C$ results upon multiplying
(\ref{eq:singleagent}) by $q(t^\prime)$ and subsequently averaging
over the noise. We make use of the identity $\bra
\eta(t)q(t^\prime)\ket_* =\sqrt{\alpha}\sum_{s}(\Sigma_{ts}G_{t^\prime
s})$ (derived via an integration by parts in the generating
functional; see \cite{HeimCool01} for an analogous identity in the
conventional MG). To deal with $G$ (which obeys causality:
$G_{tt^\prime}=0$ for $t\leq t^\prime$) one takes a field derivative
of (\ref{eq:singleagent}), followed by averaging. The result reads:
\begin{eqnarray}
\left[1+\lambda(t+1)\right]C_{t+1,t^\prime} & = & C_{tt^\prime}
 + \alpha  [(\id+G)^{-1} D
(\id+G^T)^{-1}G^T]_{tt^\prime} \nonumber\\ &&
 -
\alpha [ (\id+G)^{-1}C]_{tt^\prime},
 \label{eq:C}
\\
\left[1+\lambda(t+1)\right]G_{t+1,t^\prime} & = & G_{tt^\prime} -
\alpha [(\id+G)^{-1}G]_{t t^\prime} + \delta_{tt^\prime}.
\label{eq:G}
\end{eqnarray}
These coupled equations have to be solved subject to the constraint
$C_{tt}=r^2$ for all $t\geq 0$.  Furthermore, as in \cite{HeimCool01}
one finds for $N\to \infty$ that the rescaled disorder-averaged
average bid $\overline{\bra A_t\ket}$ is zero at any time, and that
the disorder-averaged volatility matrix (\ref{eq:volamatrix}) is
proportional to the covariance matrix of the single-trader noise:
\begin{equation}
\lim_{N\to\infty} \overline{\Xi}_{tt^\prime} =
\frac{1}{2}\Sigma_{tt^\prime}=\frac{1}{2} \left[(\id+G)^{-1} D
(\id+G^T)^{-1}\right]_{tt^\prime}. \label{eq:volamatrix2}
\end{equation}
\vsp

The dynamic order parameters in the spherical model are prescribed
in full by (\ref{eq:C},\ref{eq:G}), with the constraints
$C_{tt}=r^2$ and $1+\lambda(t)>0$. As in \cite{HeimCool01}, they
can be calculated iteratively, starting from $(t,t^\prime)=(0,0)$,
and upon using causality (i.e. $[G^n]_{tt^\prime}=0$ for
$n>t-t^\prime$). Given the prescribed values $C_{00}=r^2$ and
$G_{00}=0$ one finds, for instance:
\begin{eqnarray}
\lambda(1)&=&-1+(1+\alpha(r^{-2}\!-1)+\alpha^2)^{1/2}
\label{eq:lambda},\\
G_{10}&=&\left[1+\alpha(r^{-2}\!-1)+\alpha^2\right]^{-\frac{1}{2}}
\label{eq:G10},\\
C_{10}&=&\frac{r^2(1-\alpha)}{\sqrt{1+\alpha(r^{-2}\!-1)+\alpha^2}}
\label{eq:C10}
\end{eqnarray}
(note: $G_{11}=0$ and $C_{11}=r^2$). From these follow the
volatility matrix elements:
\begin{eqnarray}
\Sigma_{00}&=&1+r^2, \label{eq:sigma00}\\
\Sigma_{10}&=&1-\frac{\alpha
r^2+1}{\sqrt{1+\alpha(r^{-2}\!-1)+\alpha^2}}, \label{eq:sigma10}
\\ \Sigma_{11}&=&1+r^2-\frac{2}{\sqrt{1+\alpha(r^{-2}\!-1)+\alpha^2}}
+\frac{1-r^2(1-2\alpha)}{1+\alpha(r^{-2}\!-1)+\alpha^2}
\label{eq:sigma11}.
\end{eqnarray}
This iteration can be carried out for an arbitrary number of time
steps; in practice, however, the terms become prohibitively more
complicated for increasing times. Alternatively, one may iterate
(\ref{eq:C},\ref{eq:G}) numerically. Unlike the procedure proposed in
\cite{EissOppe92} to generate realisations of the single trader
process, the iteration of (\ref{eq:C},\ref{eq:G}) does not require
averaging over the single agent noise, and thus provides very precise
data (albeit that numerically inverting the $t\times t$ matrices in
(\ref{eq:C},\ref{eq:G}) becomes more and more costly as the number of
time steps increases).

One observes that, due to the explicit form of (\ref{eq:C},\ref{eq:G})
(and in sharp contrast to the similar calculation in
\cite{HeimCool01}), the temporal evolution of the macroscopic order
parameters is completely {\em independent of initial conditions}: as
long as the constraint $\lim_{N\to\infty} N^{-1}\sum_{i=1}^N
q_i(0)^2=C_{00}=r^2$ is met the distribution $P(q_i(0))$ from which
the initial point differences are drawn is irrelevant for the values
of both $C$ and $G$ at finite times as well as for the macroscopic
stationary state. We have verified this in numerical simulations,
initializing the dynamics with different distributions for the
$q_i(0)$, but all with second moment $N^{-1}\sum_{i=1}^N
q_i^2(0)=r^2$. This property of the spherical MG is quite distinct
from the conventional MG, where the explicit analysis of the first few
time steps as presented in \cite{HeimCool01} reveals that the values
of the correlation and response functions at finite times depend on
the higher moments of $P(q_i(0))$ as well, and not only on its
variance. As far as the stationary state of the conventional MG is
concerned the interest so far has mainly focussed on starts of the
form $|q_i(0)|=q_0$ for all $i$, with $q_0=0$ for {\em tabula rasa}
starts and $q_0>0$ for biased starts. Crucial differences between the
two cases have been found in the non-ergodic regime
\cite{ChalMarsZecc00,MarsChalZecc00,GarrMoroSher00}. We have extended
this analysis and have performed simulations of the conventional batch
game with different initial distributions for the $q_i(0)$ all with
the same second moment $q_0^2$. For fixed second moment we find that
{\em qualitatively} the stationary volatility does not depend on the
higher moments of $P(q_i(0))$, but that differences in the {\em
quantitative} values are found for different shapes of the initial
distribution of point differences.

Let us finally inspect the solution of (\ref{eq:C},\ref{eq:G}) for
small $\alpha$ and finite times. By induction one finds
\bd \lambda(t)={\cal
O}(\alpha),~~~~~~~~ G_{tt^\prime}=\Theta_{tt^\prime}+{\cal
O}(\alpha),~~~~~~~~ C_{tt^\prime}=r^2+{\cal O}(\alpha), \ed
 where
$\Theta_{tt^\prime}=1$ if $t>t^\prime$ and $\Theta_{tt^\prime}=0$
otherwise.  For the volatility we find $\Sigma_{tt}={\cal
O}(\alpha^2)$ for all finite $t\geq 2$. We conclude that for small
$\alpha$ the system is in a completely frozen state, with a
divergent integrated response and vanishing volatility.

\section{Analysis of Stationary States}

\subsection{Implications of Time-Translation Invariance}

We now focus on time-translation invariant solutions of the dynamical
equations (\ref{eq:C},\ref{eq:G}), i.e. we consider the system long
after any initial equilibration and study solutions of the form
\begin{equation}\label{eq:tti}
C_{tt^\prime}= C(t-t^\prime), ~~~~~~~~ G_{tt^\prime}=
G(t-t^\prime), ~~~~~~~~ \lambda(t) = \lambda.
\end{equation}
Then all matrices in (\ref{eq:C},\ref{eq:G}) become Toeplitz matrices,
and hence they commute. Given (\ref{eq:tti}) it is natural to express the
dynamics in terms of the Fourier transforms of the correlation and
response functions.  We use the following notation:
\begin{equation}
C(\tau) = \int_{-\pi}^{\pi} \frac{d\omega}{2\pi} e^{i\omega \tau}
\tilde C(\omega), ~~~~~~~~ \tilde C(\omega) = \sum_\tau
e^{-i\omega \tau} C(\tau)
\end{equation}
and similarly for $G$.  The equations (\ref{eq:C},\ref{eq:G})
subsequently translate into
\begin{eqnarray}
\tilde\Delta(\omega) \tilde C(\omega) &=&  \frac{\alpha\tilde
D(\omega)\tilde G(\omega)^*}{ |1+\tilde G(\omega)|^2} -
\frac{\alpha \tilde C(\omega)}{1+\tilde G(\omega)}, \label{eq:feqC}
\\
\tilde\Delta(\omega) \tilde G(\omega) &=& 1- \frac{\alpha \tilde
G(\omega)}{1+\tilde G(\omega)},  \label{eq:feqG}
\end{eqnarray}
where $\tilde\Delta(\omega)=(1+\lambda)e^{i\omega}-1$, and $\tilde
G(\omega)^*$ denotes the complex conjugate of $\tilde G(\omega)$.
Since $\tilde D(\omega)=\tilde C(\omega)+2\pi \delta(\omega)$, and
upon defining the integrated static response $\chi=\sum_{\tau}
G(\tau)=\tilde{G}(0)$,
 we may rewrite
(\ref{eq:feqC}) as
\begin{equation}\label{eq:fsolC}
\bigg[\tilde \Delta(\omega) |1+\tilde G(\omega)|^2 +
\alpha\bigg]\tilde C(\omega) =2\pi\alpha\chi\delta(\omega).
\end{equation}
Note that considering the case $\omega=0$ in (\ref{eq:feqG}) allows us
to express $\lambda$ in terms of $\chi$: \be \lambda =
\frac{1+\chi(1-\alpha)}{\chi(1+\chi)}
\label{eq:lambdainchi}. \ee Finally, in the stationary state the
volatility matrix $\overline\Xi$ is also of the Toeplitz form
$\overline{\Xi}_{tt^\prime}=\overline{\Xi}(t-t^\prime)$ and thus
the volatility $\sigma^2 =\overline{\Xi}(0)$ can be expressed as
\begin{equation}
\sigma^2 =\frac{1}{2}\int_{-\pi}^{\pi} \frac{d\omega}{2\pi}
\frac{\tilde C(\omega)}{|1+\tilde G(\omega)|^2}
+\frac{1}{2(1+\chi)^2}.
 \label{eq:vola_tti}
\end{equation}
Since initial conditions play no role in the macroscopic dynamics,
we must conclude that as soon as multiple stationary solutions
exist only one of these will ever be realized.
\begin{figure}[t]
\vspace*{-6mm} \hspace*{35mm} \setlength{\unitlength}{1.8mm}
\begin{picture}(100,55)
\put(-8,-0){\epsfysize=50\unitlength\epsfbox{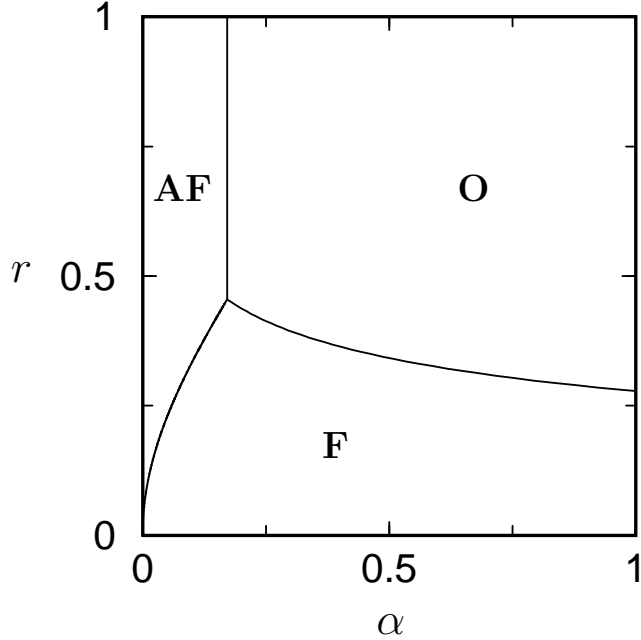}}
\put(-8,29){\Large $r$} \put( 19,3){\Large $\alpha$}
\put(15,16){\large\bf F} \put(25,35){\large\bf O}
\put(2.5,35){\large\bf AF}
\end{picture}
\vspace*{-8mm} \caption{Phase diagram of the spherical MG,
displaying three phases: {\bf O}, oscillating correlation function
and finite integrated response; {\bf F}, frozen phase with finite
integrated response; and {\bf AF}, anomalous frozen phase with
diverging integrated response. Throughout the phase AF the
volatility is zero. The O$\to$F transition is defined by
(\ref{eq:OtoF}), and the F$\to$AF transition is defined by
(\ref{eq:FtoAF}); both are continuous. The discontinuous
transition from O$\to$AF  occurs at $\alpha=3-2\sqrt{2}\approx
0.172$. The triple point corresponds to $\alpha=3-2\sqrt{2}$ and
$r=r^*\approx 0.455$.} \label{phasediagram}
\end{figure}

Before we give a detailed account of the further analysis of the
dynamical equations, we briefly summarise our results. We find that,
depending on the control parameters $\alpha$ and $r$, the system
displays three distinct phases, as illustrated in
figure \ref{phasediagram}:
\begin{enumerate}
\item a phase with finite integrated response and oscillatory
behaviour of the correlation function (O), 
\item a frozen phase with finite integrated response (F), 
\item a frozen phase exhibiting anomalous integrated response which
grows linearly with time (AF).
\end{enumerate}
We will now proceed to obtain exact solutions of the dynamical
equations in each of the three phases.

\subsection{Stationary States without Anomalous Response}

We first inspect stationary states with finite $\chi$, i.e. those for
which perturbations will decay sufficiently fast. It now follows from
(\ref{eq:fsolC}) that, for any $\omega\neq 0$, $\tilde C(\omega)$ can
be non-zero only if $\tilde \Delta(\omega) |1+\tilde G(\omega)|^2 +
\alpha=0 $. This requires $\tilde\Delta(\omega)$ to be real, which is
possible only for $\omega=0,\pi$. We conclude that $\tilde
C(\omega)=2\pi c_0 \delta(\omega)+ 2\pi c_1 \delta(\omega-\pi)$, or
equivalently
\begin{equation}
C(\tau) = c_0 + c_1 (-1)^\tau. \label{eq:Ctau}
\end{equation}
Here  $c_0$ and $c_1$ (which will depend on the parameters
$\alpha$ and $r$, as will $\chi$) are coupled via the spherical
constraint: $c_0+c_1=r^2$. Insertion of (\ref{eq:Ctau}) into
(\ref{eq:fsolC}) leads to the following two coupled equations
\be c_0\left[\alpha+\lambda(1+\chi)^2\right]=\alpha\chi,~~~~~~~~
(r^2\!-c_0)\left[\alpha-(\lambda+2)(1+\chi^\prime)^2\right]=0
\label{eq:c0eqns} \ee with $\chi^\prime=\sum_t (-1)^t
G(t)=\tilde{G}(\pi)$ measuring the response to persistent oscillating
perturbations. These equations are to be solved in combination with
(\ref{eq:feqG}). We conclude that there are two types of stationary
states with finite $\chi$: a frozen state, where $c_0=r^2$ (so
$c_1=0$), and an oscillating state, where $c_0<r$. We will work out
their properties separately below.  For solutions of the form
(\ref{eq:Ctau}) we can also work out (\ref{eq:vola_tti}) further:
\begin{equation}\label{eq:sigmatheory}
\sigma^2=
\frac{1+c_0}{2(1+\chi)^2}+\frac{(r^2-c_0)}{2(1+\chi^\prime)^2}.
\end{equation}

\subsubsection*{Oscillating stationary states without anomalous
response.}

Here $c_0<r^2$, and the remaining four (coupled but closed)
equations to be solved to find the stationary state include one
expression for $\chi^\prime$ which one obtains by choosing
$\omega=\pi$ in (\ref{eq:feqG}):
\begin{eqnarray}
c_0=\frac{\alpha\chi}{\alpha+\lambda(1+\chi)^2}, &~~~~~~~~&
\alpha=(\lambda+2)(1+\chi^\prime)^2, \label{eq:seta}
\\ \lambda =
\frac{1+\chi(1-\alpha)}{\chi(1+\chi)},&~~~~~~~~& \lambda+2 =-
\frac{1+\chi^\prime(1-\alpha)}{\chi^\prime(1+\chi^\prime)}.
\label{eq:setb}
\end{eqnarray}
The set (\ref{eq:seta},\ref{eq:setb}) allows for two types of
solutions. The first, where $\lambda=\alpha-1+2\sqrt{\alpha}$,
obeys the requirement $1+\lambda>0$ for all $\alpha$. One must in
fact demand $\lambda>0$ in order to have a finite $\chi$ (as
required), and we reject $\chi<0$ solutions on physical grounds.
This leaves:
\begin{eqnarray}
\lambda&=&\alpha-1+2\sqrt{\alpha}, \label{eq:lambda_O}\\ \chi&=&
\frac{1-\alpha-\sqrt{\alpha}+
\sqrt{2\alpha^{3/2}+\alpha^2}}{-1+2\sqrt{\alpha}+\alpha},
\label{eq:chi_O}
\\
\chi^\prime&=&  - \frac{1}{1+\sqrt{\alpha}}.
\label{eq:chiprime_O}
\end{eqnarray}
 We note that $\lambda$, $\chi$, $\chi^\prime$ and $c_0$ are
independent of $r$; only $c_1$ depends on $r$ via $c_1=r^2-c_0$. The
second type of solution, where $\lambda=\alpha-1-2\sqrt{\alpha}$,
meets our requirement $\lambda+1>0$ only for $\alpha>4$. It turns out
that such solutions are never realized, so we will not give their
equations in full. We have now determined all order parameters and the
volatility in explicit form: $c_0$ follows from insertion of
(\ref{eq:lambda_O},\ref{eq:chi_O}) into the first equation of
(\ref{eq:seta}), whereas the volatility follows upon inserting $c_0$
and (\ref{eq:chi_O},\ref{eq:chiprime_O}) into
(\ref{eq:sigmatheory}). For $\alpha\to\infty$ one finds
$\lim_{\alpha\to\infty}\chi=0$,
$\lim_{\alpha\to\infty}\lambda/\alpha=1$, so that
$\lim_{\alpha\to\infty}c_0=0$; the amplitude of the oscillations in
the correlations increases with increasing $\alpha$.

The present solution breaks down when either $\chi\to \infty$ or
 $c_1\to 0$. The corresponding mathematical conditions are found to be
$\alpha=\alpha_{c,1}$ (with $\chi<\infty$ for
$\alpha>\alpha_{c,1}$) and $\alpha=\alpha_{c,2}(r)$ (with $c_1>0$
for $\alpha>\alpha_{c,2}(r)$), respectively, where
\begin{eqnarray}
\alpha_{c,1}&=&3-2\sqrt{2}~\approx ~0.172, \label{eq:OtoAF}
\\
\alpha_{c,2}(r)&=&\left[1-\frac{2+1/r^2}{2\sqrt{1+1/r^2}}\right]^2.
\label{eq:OtoF}
\end{eqnarray}
We note that $\alpha_{c,1}=\alpha_{c,2}(r)$ at
$r=r^*=\sqrt{\alpha_{c,1}/(1-\alpha_{c,1})}\approx 0.455$, and that
$\alpha_{c,2}(r)>\alpha_{c,1}$ for $r<r^*$ and
$\alpha_{c,2}(r)<\alpha_{c,1}$ for $r>r^*$.  Thus one expects that, as
$\alpha$ is lowered for any $r>r^*$, the amplitude of the oscillations
of the correlations remains positive until the critical value
$\alpha_{c,1}$ is reached and a transition to a state with anomalous
response occurs. At this point one has \be \lim_{\alpha\downarrow
\alpha_{c,1}}
\sigma^2=\frac{1}{3-2\sqrt{2}}\left(r^2-\frac{3-2\sqrt{2}}{2(\sqrt{2}-1)}\right).
\label{eq:limitsigma} \ee
 For $r<r^*$ the oscillatory behaviour of the correlations breaks down
as $\alpha$ is lowered before anomalous response sets in, and the
system enters a frozen state with finite integrated response.

\subsubsection*{Frozen stationary states without anomalous response.}

Here $c_0=r^2$, and there is no need to calculate $\chi^\prime$;
the coupled equations to be solved are simply
\begin{eqnarray}
\lambda\chi^2+(2\lambda-\alpha/r^2)\chi+\lambda+\alpha &=& 0, \\
\lambda\chi^2+(\alpha-1+\lambda)\chi-1 &=& 0,
\end{eqnarray}
from which we can determine $\lambda$ and $\chi$ to be:
\begin{eqnarray}
\lambda&=&-1-\alpha+\frac{(2+1/r^2)\sqrt{\alpha}}{\sqrt{1+1/r^2}},
\label{eq:lambda_F}\\
\chi&=&[\sqrt{\alpha}\sqrt{1+1/r^2}-1]^{-1}\label{eq:chi_F}.
\end{eqnarray}
Upon inserting (\ref{eq:chi_F}) and the relation $c_0=r^2$, one
finds that our exact expression (\ref{eq:sigmatheory}) for the
volatility simplifies to
\begin{equation}
\sigma^2= \frac{1}{2}\left[\sqrt{r^2+1}-r/\sqrt{\alpha}\right]^2.
\label{eq:vola_F}
\end{equation}
The present frozen state will cease to be a consistent solution at
the point where $\chi$ diverges. Equation (\ref{eq:chi_F}) states
that this happens at $\alpha=\alpha_{c,3}(r)$, where
\begin{equation}\label{eq:FtoAF}
\alpha_{c,3}(r)=r^2/(r^2+1).
\end{equation}
The line $\alpha=\alpha_{c,3}(r)$ marks the transition between a
frozen state with finite integrated response (for
$\alpha>\alpha_{c,3}(r)$), and an anomalous frozen state (for
$\alpha<\alpha_{c,3}(r)$). According to (\ref{eq:vola_F}) it also
coincides with the line where the volatility vanishes.

\subsection{Frozen States with Anomalous Response}

Numerical simulations for small $\alpha$ and large $r$ reveal a
parameter regime with a stationary state in which the volatility
vanishes, $\sigma^2=0$, and where all agents are frozen in such a way
that \be \label{eq:AF} C(\tau)=r^2,~~~~~~~~
\lambda(\tau)=0,~~~~~~~~~\chi=\infty. \ee These three identities
clearly hold at the transition line (\ref{eq:FtoAF}), where anomalous
response first emerges in the frozen state. In this subsection we
demonstrate that our dynamical equations (\ref{eq:C},\ref{eq:G})
indeed allow for self-consistent stationary state solutions with the
properties (\ref{eq:AF}).  Note that (\ref{eq:AF}) directly imply that
$\sum_\tau [\id+G]^{-1}(\tau)=(1+\chi)^{-1}=0$, and hence also
$\sigma^2=0$.  Insertion of (\ref{eq:AF}) as {\em ans\"{a}tze} into
(\ref{eq:C},\ref{eq:G}) now gives:
\begin{eqnarray}
0 & = &  [(\id+G)^{-1} D (\id+G^T)^{-1}G^T](\tau) - [
(\id+G)^{-1}C](\tau), \label{eq:simpleC}
\\
G(\tau+1) & = & G(\tau) + \alpha (\id+G)^{-1}(\tau) +
(1-\alpha)\delta(\tau), \label{eq:simpleG}
\end{eqnarray}
where $\delta(\tau)=1$ if $\tau=0$ and $\delta(\tau)=0$ otherwise.
We define the persistent response
$g=\lim_{T\to\infty}T^{-1}\sum_{\tau\leq T}G(\tau)$, and sum both
sides of  (\ref{eq:simpleG}) from $\tau=0$ to $\tau=\ell$, to
get:
\be
 G(\ell+1)  =  1-\alpha+ \alpha \sum_{\tau\leq \ell}(\id+G)^{-1}(\tau).
 \ee
 We conclude that $g=1-\alpha$. From this, in turn, one infers for $\alpha<1$ that
indeed $\chi=\sum_\tau G(\tau)=\infty$, which confirms in
retrospect that our {\em ans\"{a}tze} (\ref{eq:AF}) indeed solve
our dynamical laws (\ref{eq:simpleC},\ref{eq:simpleG}).

\section{The Phase Diagram and its Verification}

\begin{figure}[t]
\vspace*{1mm} \hspace*{4mm}
\begin{tabular}{cc}
\epsfxsize=70mm \epsffile{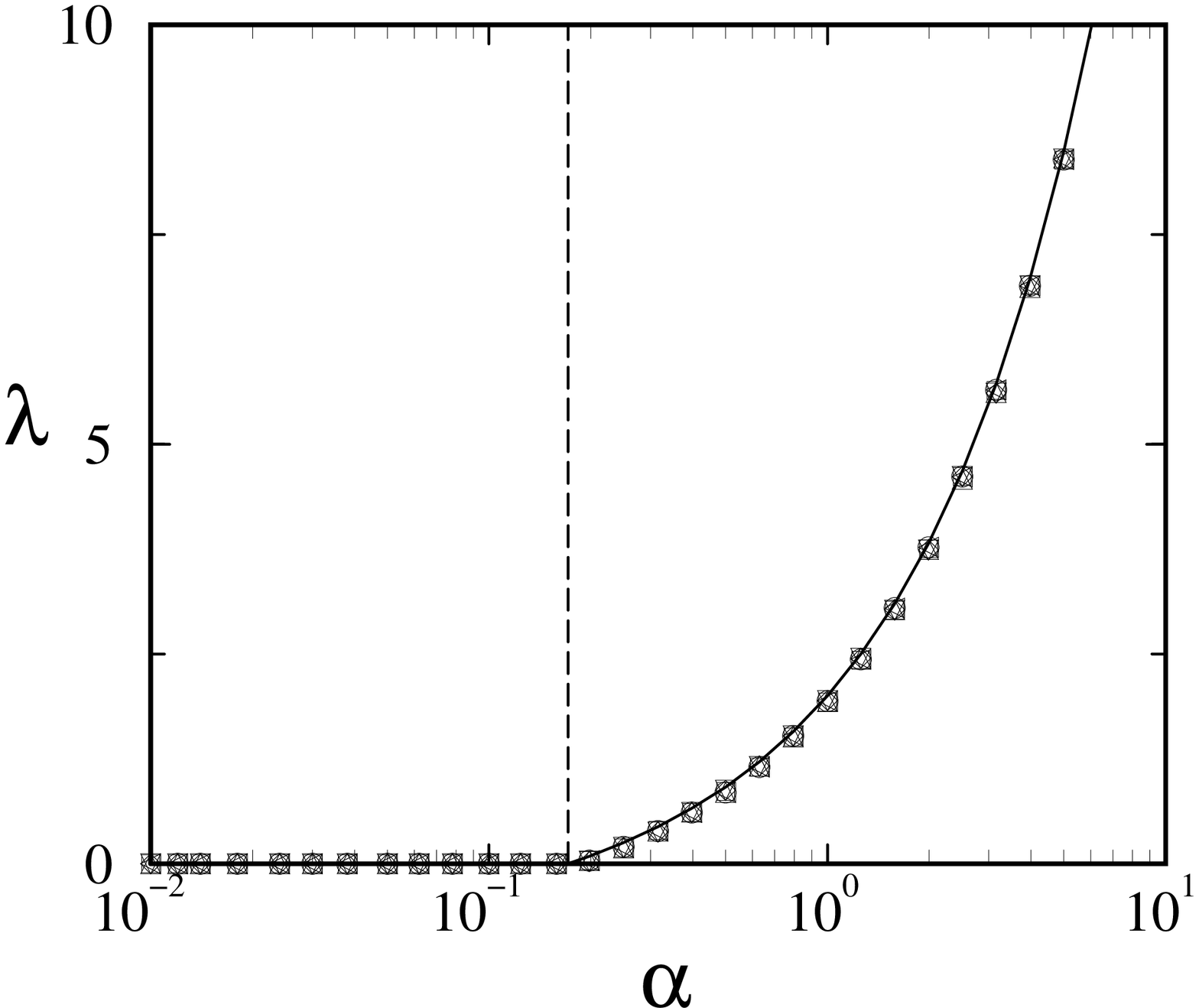} &
~~\epsfxsize=70mm \epsffile{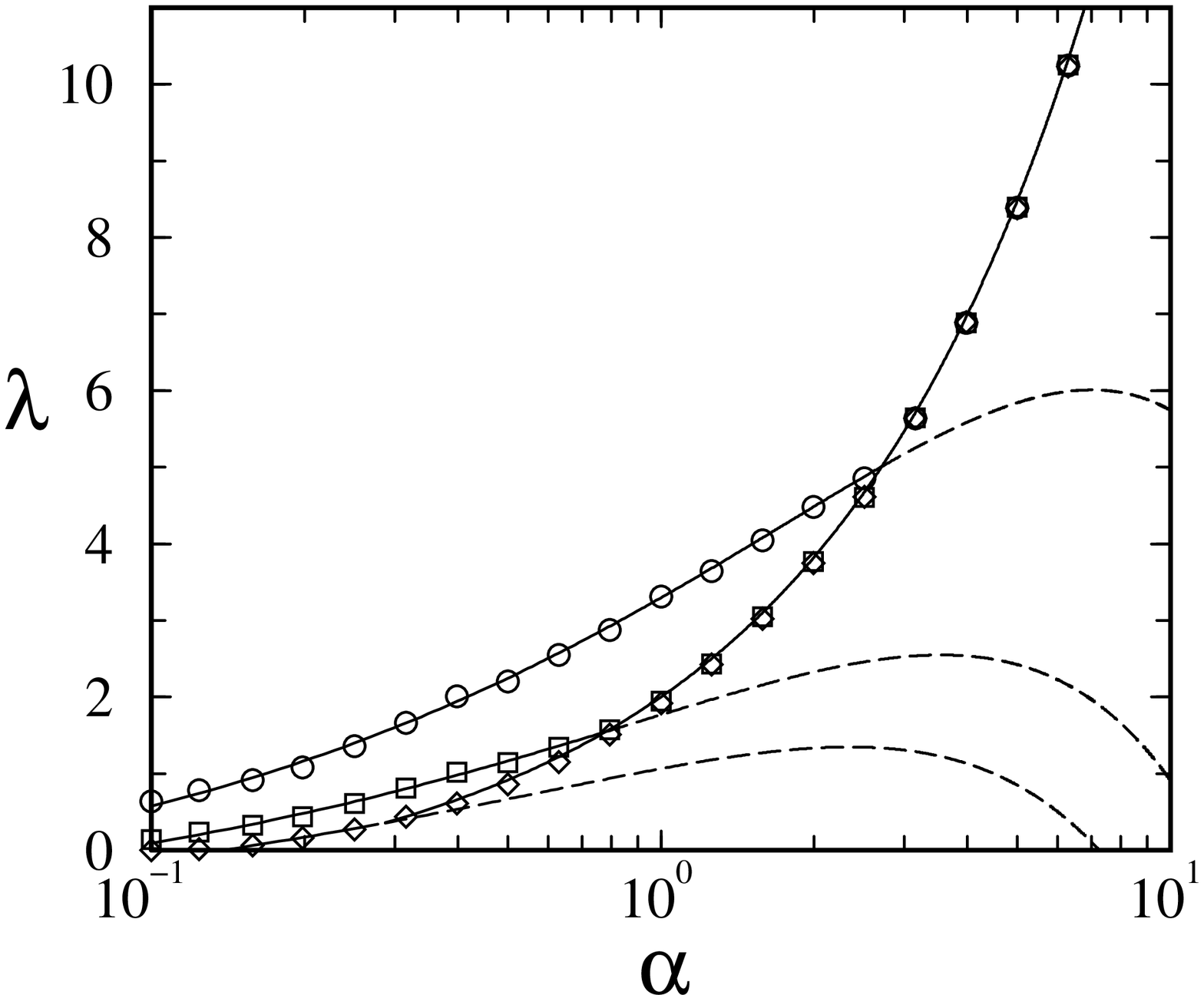}
\\[5mm]
\hspace*{-2mm} \epsfxsize=73mm \epsffile{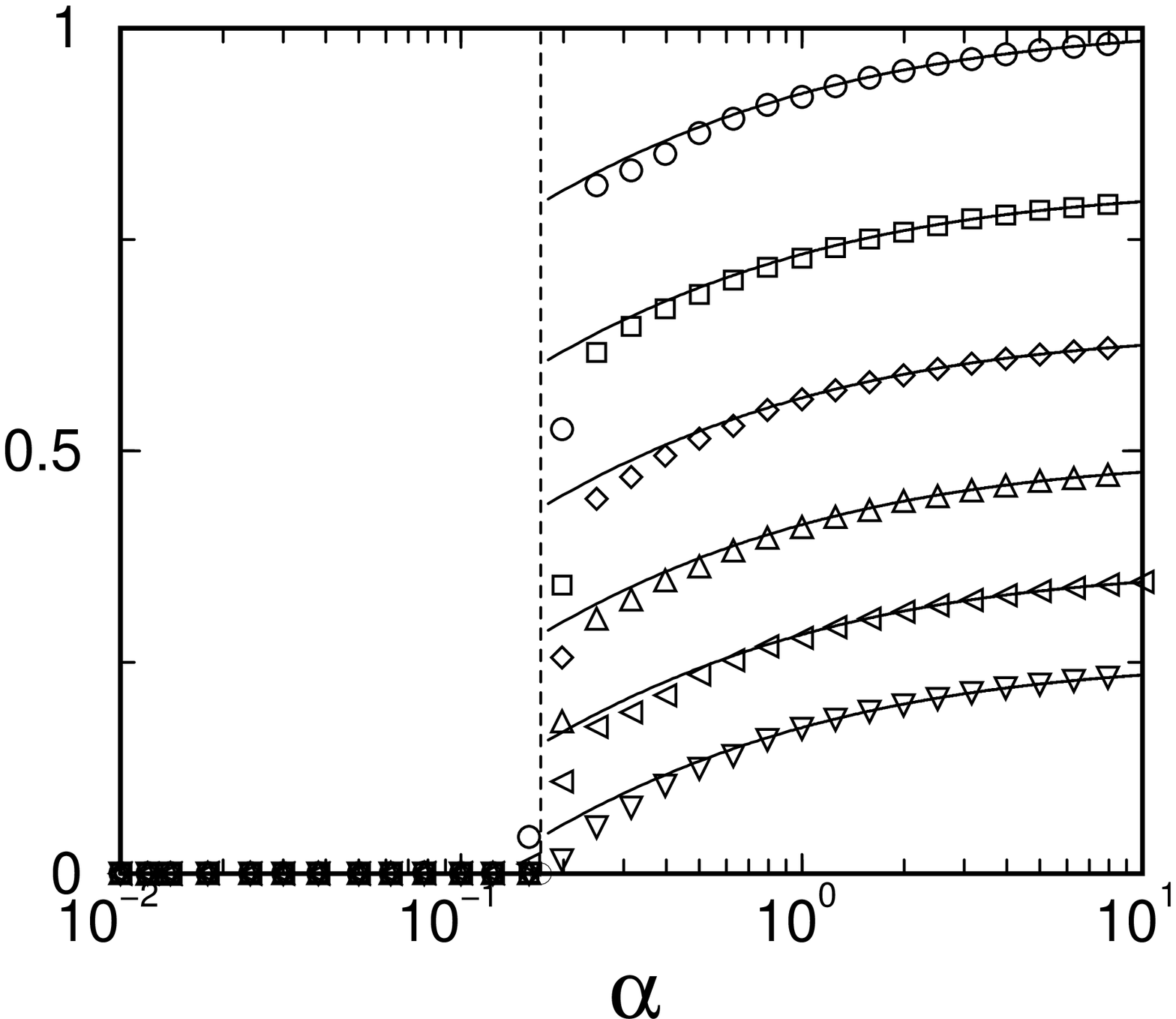}~ &
~\epsfxsize=73mm \epsffile{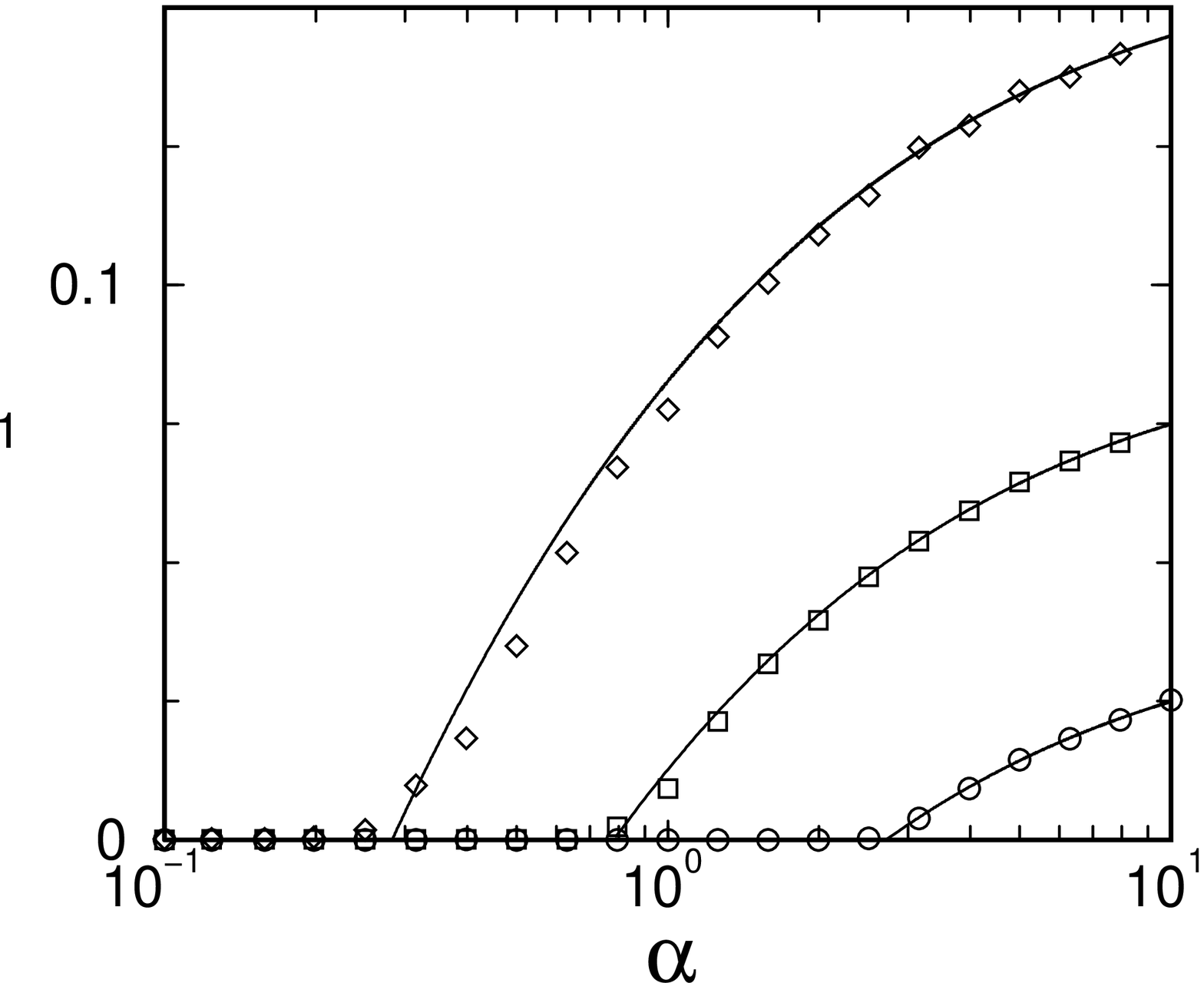}
\\[5mm]
\hspace*{-2mm} \epsfxsize=73mm \epsffile{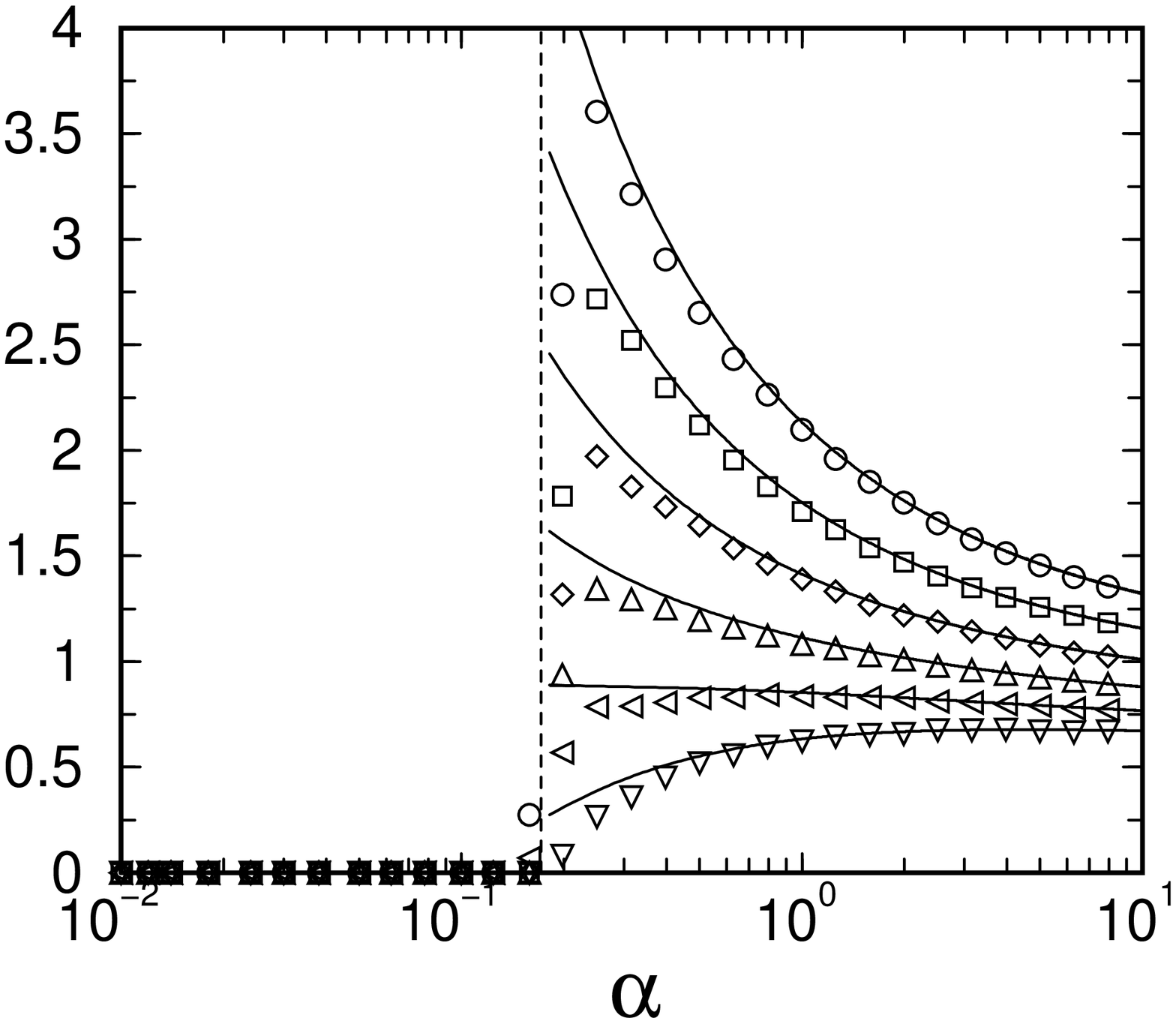}~ &
~\epsfxsize=73mm \epsffile{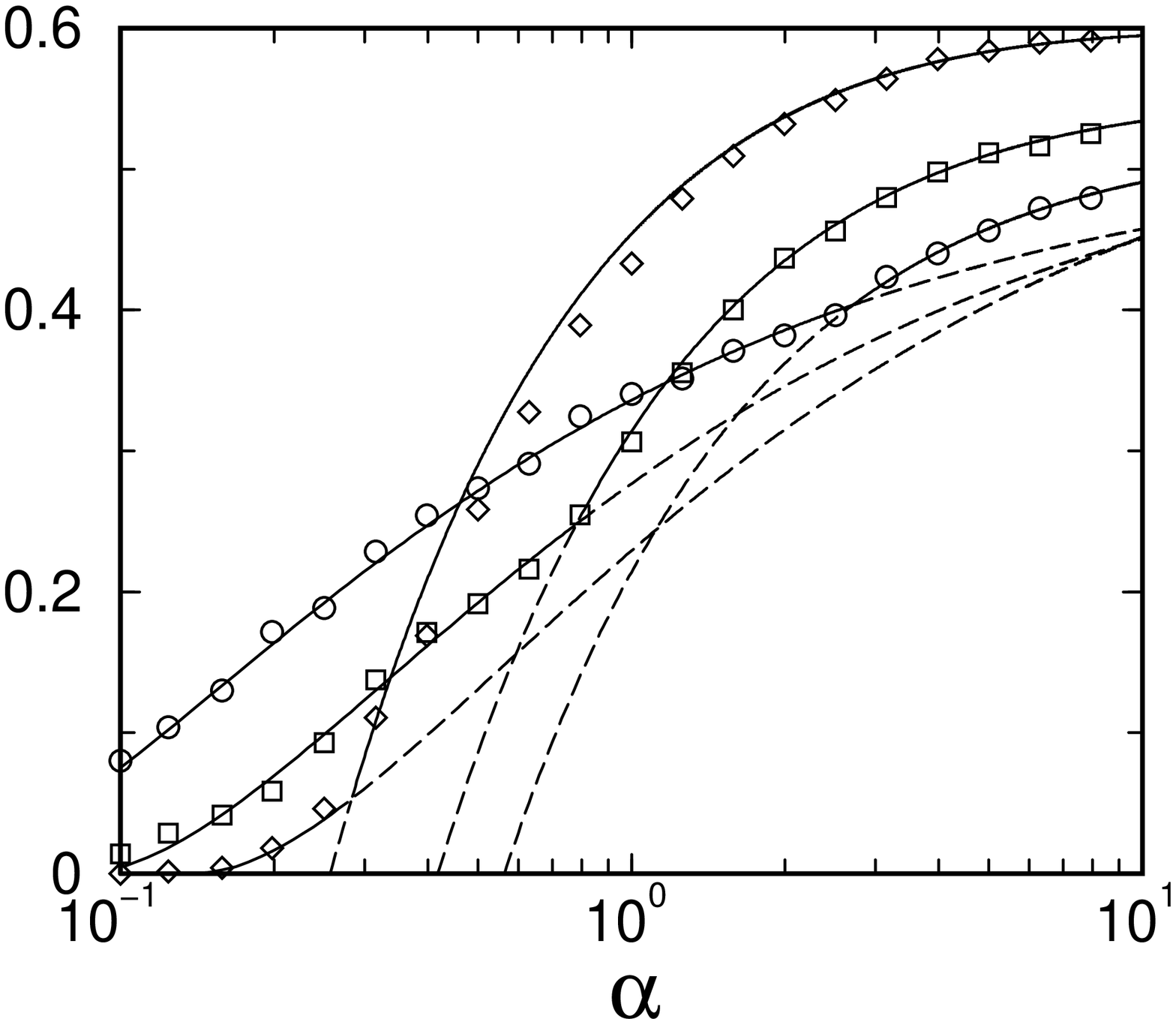}
\end{tabular}
\vspace*{3mm} \caption{Verification of O$\to$AF and O$\to$F
transitions. Left pictures: Lagrange parameter $\lambda$, oscillation
amplitude $c_1$ and volatility $\sigma^2$ versus $\alpha$, for $r=1.0,
0.9, 0.8, 0.7, 0.6, 0.5$ (from top to bottom in the panels showing
$c_1$ and $\sigma^2$). Here we expect to see the O$\to$AF
transition. Markers: simulations; solid lines: theory. The dashed
vertical line marks the predicted location $\alpha_{c,1}$ of the
O$\to$AF transition. Right pictures: the same observables shown versus
$\alpha$, but now for $r=0.4, 0.3, 0.2$; here we expect to see the
O$\to$F transition.  Markers: simulations for $r=0.2$ (circles),
$r=0.3$ (squares) and $r=0.4$ (diamonds); solid lines: theory.  The
lines are continued as dashed lines into the opposite phases, where
they should no longer be valid. }
\label{fig:figure2} \vspace*{-12mm}
\end{figure}

In the previous section we have identified three distinct phases,
which according to the extensive numerical simulations described below
exhaust the phase diagram in the ($\alpha,r)$-plane of the present
spherical MG. We have also derived explicit expressions for the
macroscopic order parameters and the volatility in all three phases,
and we have been able to calculate the various phase boundaries in
explicit form. Our results may be summarised as follows: \bd
\begin{array}{ll}
\alpha>\max\left\{\alpha_{c,1},~\alpha_{c,2}(r)\right\}: & {\rm
oscillating~ phase~ (O)}\\ & {\rm oscillating}~ C(t),~{\rm
finite}~\chi\\[2mm]  r<r^*,\,\alpha_{c,3}(r)<\alpha<\alpha_{c,2}(r): &
 {\rm frozen~ phase~ (F)}\\ & {\rm constant}~ C(t),~ {\rm
 finite}~\chi\\[2mm]
\alpha<\min\left\{\alpha_{c,1},~\alpha_{c,3}(r)\right\}: & {\rm
anomalous~ frozen~ phase~ (AF)}\\ & {\rm constant}~ C(t),~{\rm
infinite}~\chi
\end{array}
\ed
with
\be
\hspace*{-10mm} \alpha_{c,1}=3-2\sqrt{2}, ~~~~~~~~
\alpha_{c,2}(r)=\left[1-\frac{r+1/2r}{\sqrt{r^2\!+1}}\right]^2,
~~~~~~~~\alpha_{c,3}(r)=\frac{r^2}{r^2\!+1}.
\label{eq:alltransitions} \ee The resulting phase diagram is shown in figure \ref{phasediagram}.
Let us briefly discuss its main features.  For $r>r^*$ the behaviour
of the spherical MG is similar to that of its conventional
counterparts, with a divergence of $\chi$ at some fixed critical
$\alpha$ and $\chi$ remaining infinite as $\alpha$ is reduced further
to zero. There are crucial differences though: Firstly, in the
conventional MG persistent oscillations are found only for
$\alpha<\alpha_c$, while they decay above the transition. The opposite
is the case in the spherical model. Secondly, in the conventional MG
the volatility $\sigma^2$ is a smooth function of $\alpha$ across the
transition. In the spherical model we find that the volatility (and
the amplitude $c_1$ of the oscillations as well) exhibits a jump at
$\alpha=\alpha_{c,1}$ for $r>r^*$. The discontinuity of $\sigma^2$
follows immediately from the nonzero value of (\ref{eq:limitsigma})
for $r>r^*$, which gives $\sigma^2$ in the phase O close to the
O$\to$AF transition, whereas one has $\sigma^2=0$ throughout the AF
phase. The magnitude of the jump decreases as $r$ is lowered and
finally vanishes at $r=r^*$.  Below $r=r^*$ no discontinuities are
present and one finds an intermediate regime, where the system
freezes, but as yet with a finite $\chi$. Only as $\alpha$ is lowered
further a transition to a frozen phase with anomalous integrated
response takes place at $\alpha=\alpha_{c,3}(r)$, and below
$\alpha_{c,3}(r)$ both the volatility and the normalisation factor
$\lambda$ vanish identically.

We have tested our theoretical predictions against numerical
simulations of the spherical MG. The data shown in the figures are all
obtained from simulations of the batch process
(\ref{eq:sphericalupdate},\ref{eq:constraint}) with $N=500$ players,
and averaged over $20$ realisations of the disorder (i.e. the
realisations of the strategies). All measurements are temporal
averages over $250$ time steps, preceded by $250$ `equilibration'
steps. We focus on the parameter regions where the various phase
transitions are predicted to occur and depict the values of the
stationary order parameters $\lambda$ and $c_1$ as well as the
volatility $\sigma^2$ as indicators for the predicted
transitions. The precise locations of the various transitions are given in
(\ref{eq:alltransitions}).

\begin{figure}[t]
\vspace*{1mm}
\begin{tabular}{cc}
\epsfxsize=72mm  \epsffile{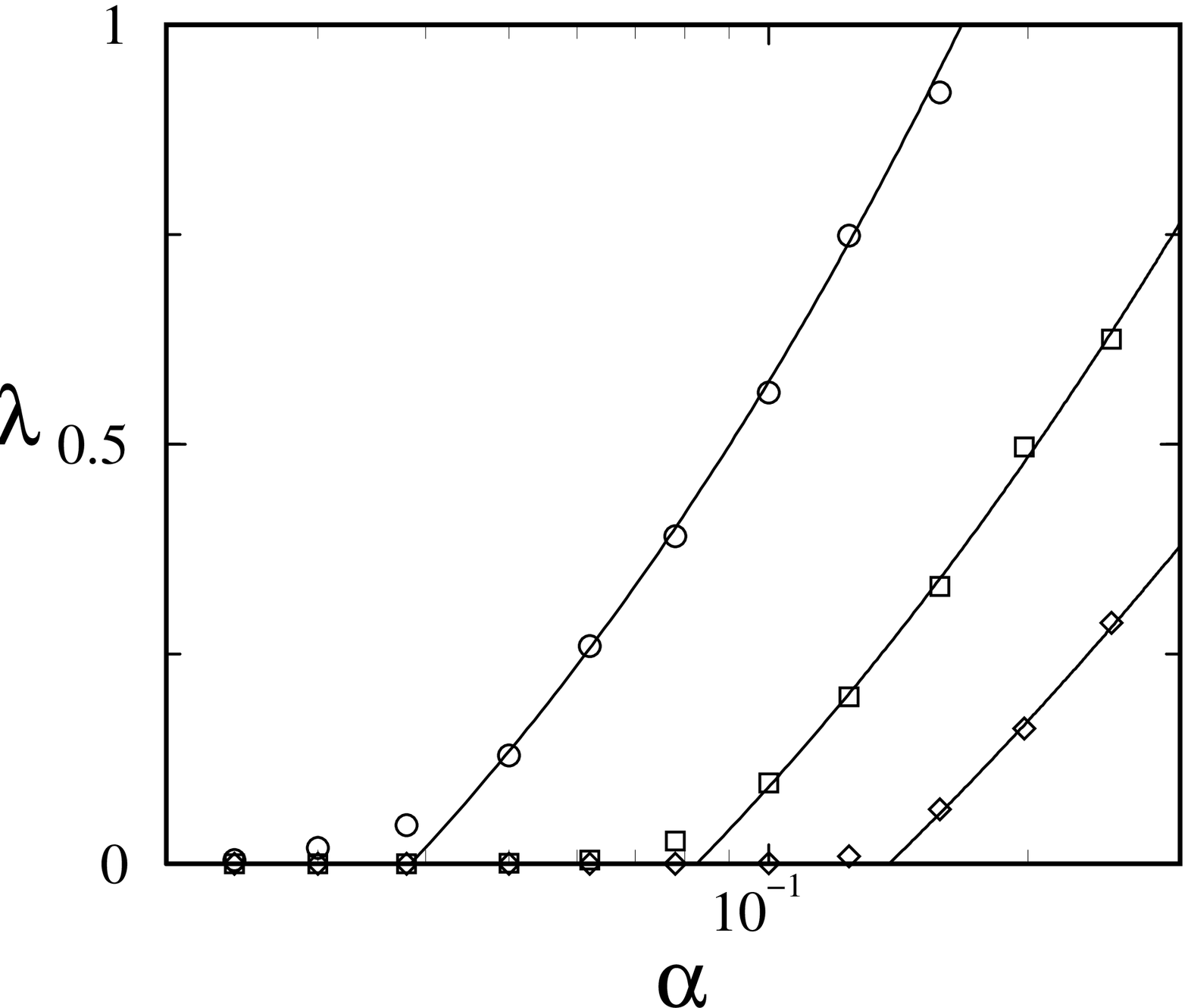} ~&~~
\epsfxsize=72mm  \epsffile{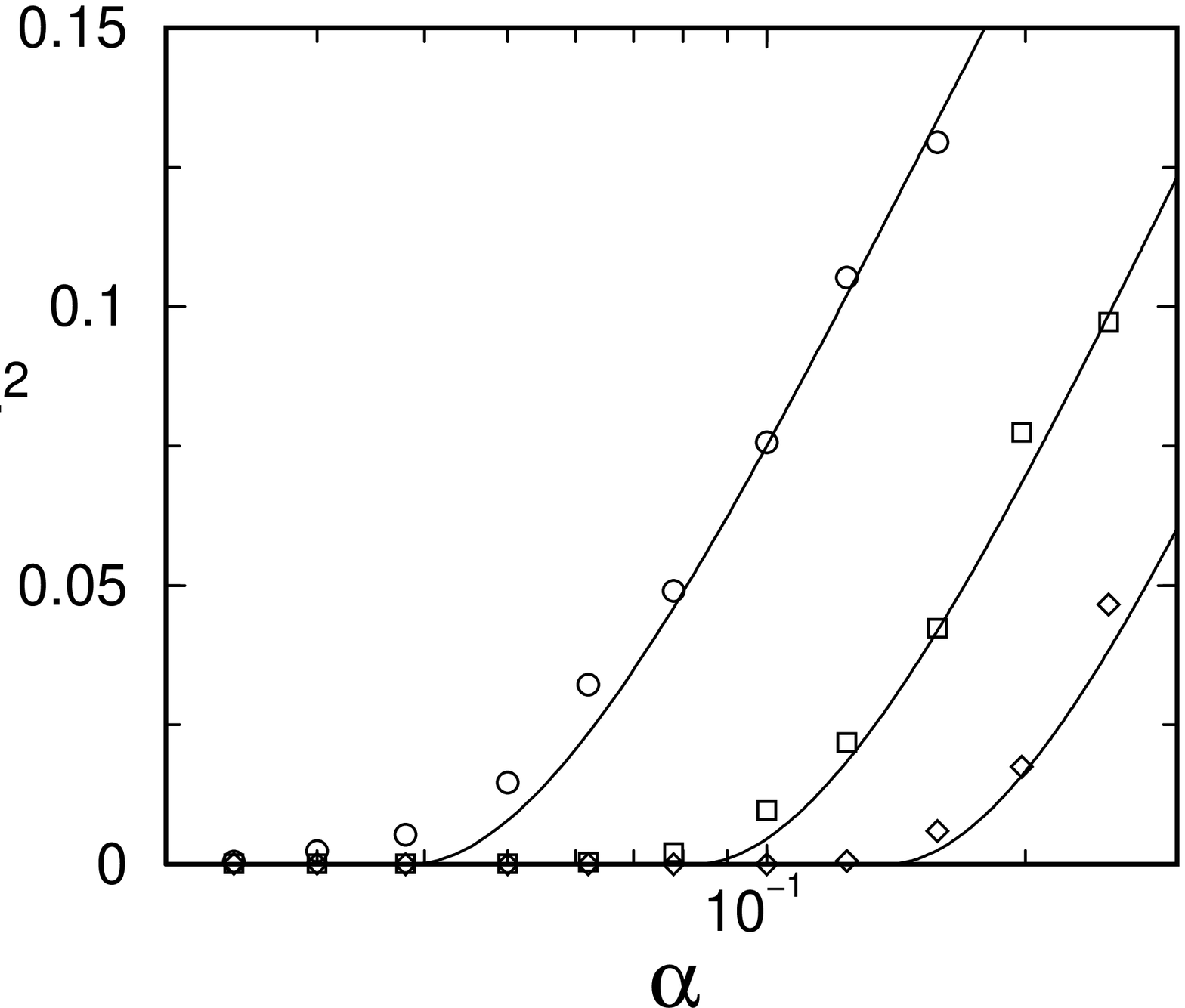}
\end{tabular}
\vspace*{4mm} \caption{ Verification  of F$\to$AF transition.
Lagrange parameter $\lambda$ (left) and volatility $\sigma^2$
(right) versus $\alpha$,
 for $r=0.2$ (circles), $r=0.3$ (squares) and $r=0.4$ (diamonds).
Solid lines: theory. Both observables are
predicted to converge to zero as the transition
$\alpha=\alpha_{c,3}(r)$ is approached from
above, and to remain zero below the transition.} \label{fig:figure3}
\end{figure}

Figure \ref{fig:figure2} concerns the O$\to$AF and O$\to$F
transitions. For $r>r^*\approx 0.455$ we expect to see the O$\to$AF
transition. Here our theory predicts that
$\lambda=\alpha-1+2\sqrt{\alpha}$ for $\alpha>\alpha_{c,1}$, and
$\lambda=0$ for $\alpha<\alpha_{c,1}$. At $\alpha_{c,1}$ the
volatility and the oscillation amplitude $c_1$ should both jump
discontinuously to zero. For $r<r^*$, on the other hand we should
observe the O$\to$F transition where $c_1$ goes to zero continuously
at $\alpha=\alpha_{c,2}(r)$. We also expect $\lambda$ and $\sigma^2$
to be continuous at this transition, albeit their derivatives with
respect to $\alpha$ change discontinuously. The data in figure
\ref{fig:figure2} reveal full agreement between theory and simulation
(up to finite size effects close to the transitions). Although we
restrict ourselves to $r\le 1$ in figure \ref{fig:figure2}, we have
verified that the qualitative behaviour of the system remains
unchanged for larger values of $r$ and that the very good agreement
between theory and simulation continues to hold for $r>1$. Figure
\ref{fig:figure3} concerns the F$\to$AF transition, where $\lambda$
and $\sigma^2$ are predicted to vanish as $\alpha$ approaches
$\alpha_{c,3}(r)$ from above. Again we find good agreement between
theory and numerical experiment.

\section{Conclusions}

In this paper we have introduced a spherical version of the batch
minority game, with random public information. In this model the
non-linear update rule of the conventional game is replaced by an
iteration prescription which is linear in the microscopic degrees of
freedom (the point differences $q_i(t)$), complemented by a spherical
constraint $N^{-1}\sum_i q_i^2(t)=r^2$. The spherical MG is designed
to be exactly solvable in the thermodynamic limit. In terms of the
decision making of the individual agents, the linearised microscopic
dynamical laws corresponds to allowing the agents to play linear
combinations of their two strategies. The relevant control parameters of
the spherical MG are the radius $r$ of the sphere to which the
dynamics is confined and the ratio $\alpha=p/N$ of the number of
possible values of the external information over the number of agents.

Using the dynamic mean field theory introduced by De Dominicis we are
able to perform the average over the disorder and to take the
thermodynamic limit. This formalism reduces the original $N$-agent
dynamics to a non-Markovian effective single-agent stochastic
process. Like in spherical spin-glass models, the temporal evolution
of the macroscopic order parameters (the correlation and response
functions) can be formulated in terms of a pair of coupled iterative
equations, without referring to the microscopic single effective-agent
process. Assuming the existence of a time-translation invariant
stationary state we are able to solve these equations exactly, and to
compute the order parameters in the stationary state as well as the
stationary volatility at every point of the phase diagram without
making any approximations.

We find that, although the update rule is relatively simple compared
to the conventional MG, the spherical MG displays a remarkably rich
structure. Depending on $r$ and $\alpha$ the system exhibits three
distinct phases, two without anomalous response (an oscillating and a
frozen state) and a further frozen phase with diverging integrated
response. As described above the spherical model exhibits some
similarities as well as intriguing differences compared to
conventional MGs. The four main differences are (i) the absence of any
macroscopic dynamical effect of the choice of the initial microscopic
state in the spherical game, (ii) the fact that, for any $r$, the
volatility is always zero close to $\alpha=0$ whereas in the
conventional MGs both high-volatility and low-volatility solutions can
be found, (iii) persistent oscillations in the spherical MG, which
increase for increasing $\alpha$ and vanish for low $\alpha$, where in
the conventional batch MG persistent oscillations can only be found in
the low-$\alpha$ regime and (iv) the discontinuous dependence of the
volatility on $\alpha$ in the spherical MG for $r>r^*$.

In summary, our study demonstrates that the dynamical rules of the
conventional MG can be simplified to obtain a completely solvable
spherical version, which still displays a non-trivial phase diagram
and some novel features, which are not observed in conventional
MGs. It would be interesting to study further the mathematical
properties of the spherical model, such as the relaxation
towards the stationary state and possible ageing phenomena.

\section*{Acknowledgements}

The authors would like to thank EPSRC for financial support under
research grant GR/M04426 and studentship 00309273. T.G. acknowledges the
award of a Rhodes Scholarship and support by Balliol College,
Oxford. Fruitful discussions with J.P. Garrahan are gratefully
acknowledged.

\end{document}